\documentclass[aps,prb,showpacs,twocolumn]{revtex4-1}
\usepackage[utf8x]{inputenc}
\usepackage{amssymb}
\usepackage{graphicx}
\usepackage{color}
\begin{document}
\title{Field-Induced Spin-Exciton Condensation 
in $d_{x^2-y^2}$-wave Superconducting CeCoIn$_5$}
\author{V. P. Michal and V. P. Mineev}
\affiliation{Commissariat \`a l'Energie Atomique,
INAC/SPSMS, 38054 Grenoble, France}
\date{\today}

\begin{abstract}
The implications of the spin exciton mechanism are exposed in the context of a Spin Density Wave (SDW) instability occurring inside the superconducting phase of a layered heavy electron compound. In this model a magnetic field serves as a tuning parameter bringing the system to the point where the \emph{transverse} spin correlations are enhanced due to d$_{x^2-y^2}$-wave superconductivity and induces an instability to a phase with coexisting superconductivity and SDW order. The model considers electrons in a crystal with antiferromagnetic interactions and provides restrictions on the Fermi surface characteristics and on the ordering wave-vector (that can be commensurate or incommensurate close to commensuration). The applications of the model are addressed to the low-temperature/high-magnetic-field phase of CeCoIn$_5$ [M. Kenzelmann \emph{et al.}, Science \textbf{321}, 1652 (2008), Phys. Rev. Lett. \textbf{104}, 127001 (2010)]\cite{Kenzelmann}. 
\end{abstract}

\maketitle
A significant experimental finding was achieved by Kenzelmann and collaborators \cite{Kenzelmann} while using Elastic Neutron Scattering to probe magnetism in the superconducting phase of CeCoIn$_5$ at low temperature and under a strong magnetic field applied in the basal plane of its tetragonal crystal structure. The upper critical field in CeCoIn$_5$ is mostly determined by paramagnetic limiting ($H_{c2}(T=0)\simeq11.7\,\text{T}$) and due to this the phase transition to the superconducting state below $T=0.4T_c$ ($T_c=2.3\,\text{K}$)\cite{Petrovic} 
is of the first order\cite{FirstOrder}. 
In paper\cite{Kenzelmann} the authors observed a magnetic ordering with wave-vector $\mathbf{Q}=(q,q,1/2)$ where $q\simeq0.45$ is an incommensurate wave vector (here we use Reciprocal Lattice Units in which $2\pi/a=1$, $a$ being the in-plane lattice constant), which was independent of the field magnitude. The small value of the magnetic moment on Cerium sites $m=0.15\mu_B$ ($\mu_B$ is the Bohr magneton) oriented along the c-tetragonal axis indicated that the ordering is of itinerant origin. It was remarkable there that incommensurate SDW was confined inside the superconducting phase, meaning that here superconductivity is an essential ingredient for SDW to develop. The existence of the magnetic order was first detected by the technique of NMR\cite{Young} and its precise field-dependence later determined\cite{Koutroulakis}.

Theoretically, these coexisting orders were discussed within models that can be divided into three classes. In the first one\cite{Kenzelmann,Agterberg,Yanase,Miyake,Aperis} theories rely on coupling between SDW, superconductivity and superconductivity with Cooper pairs having non-zero center of mass momentum (Fulde-Ferrel-Larkin-Ovchinnikov (FFLO) phase, Pair Density Wave (PDW), or $\pi$-triplet superconductivity), which stabilizes SDW and superconductivity at high-field and low-temperature. Evidence for the occurence of a staggered superconducting state in CeCoIn$_5$ is however still to be revealed. A second point of view\cite{Suzuki} highlighted the role played by the vortex lattice which can increase the density of states in the nodal direction of the gap and trigger a magnetic instability. In the third \cite{Ikeda,Kato} it was emphasized  the importance of Pauli limiting in d-wave superconductors for stabilizing SDW order in the case where the ordering wave-vector is the nesting wave-vector $Q_N$ (in the sense that it joins two points of the normal-state Fermi line where the Fermi velocities are parallel). There was found that low-temperature/high-field superconducting phase in CeCoIn$_5$ is a coexisting phase of FFLO and incommensurate SDW orders.\cite{Ikeda} 
Also, there was pointed out \cite{Kato} the enhancement of nesting by superconductivity in the gap nodal direction.

Another important observation was made on CeCoIn$_5$ thanks to Inelastic Neutron Scattering (INS). Stock and collaborators \cite{Stock} measured a spin resonance that was sharp in energy ($\omega=0.60\pm0.03\,\text{meV}$) and having wave vector distribution centered on $\mathbf{Q}=(1/2,1/2,1/2)$ with a width of $\simeq 0.15$. Thereafter Panarin and collaborators \cite{Panarin} studied the evolution of the resonance in a magnetic field applied in the [1,-1,0] direction. They observed the same resonance with a decrease in its energy and a broadening in its lineshape as the field increased. They were able to measure it up to $\simeq0.5H_{c2}$ where the signal was lost in the incoherent part of the spectrum.

Theoretically Eremin and collaborators\cite{Eremin} have attributed the resonance to the proximity to the threshold of the particle-hole excitations continuum which is at energy $\omega_c=\min(|\Delta_{\bf k}|+|\Delta_{{\bf k}+{\bf q}}|)$. Another scenario related to a magnon excitation was proposed by Chubukov and Gor'kov\cite{Chubukov}. 

In this paper we present a new direction in the interpretation of the occurence of a phase with coexisting superconductivity and SDW in CeCoIn$_5$ where the phenomena \cite{Kenzelmann} and \cite{Stock,Panarin} are closely connected: the resonance that exists at $\omega=0.6\,\text{meV}$ shifts to lower energies as a transverse magnetic field is applied and triggers a magnetic instability before superconductivity is suppressed. We consider here a situation with electrons in a crystal having antiferromagnetic correlations as was previously discussed \cite{Stock,Eremin}.

We use a model of a two-dimensional system and show that it provides a consistent scenario for the presence in CeCoIn$_5$ of a SDW order that is confined inside the superconducting state without requiring the coupling with another state like FFLO or PDW, or nesting properties of the Fermi surface. The applicability of a two-dimensional model for CeCoIn$_5$ was already discussed\cite{Chubukov} because of a lack of strict two dimensionality for the compound. Here we argue that the Fermi surface in CeCoIn$_5$ has sufficient 2D character so that we can consider a model of a 2D metal from which superconductivity develops. This assumption is corroborated by experimental observations\cite{Murphy, Settai}. In particular, \cite{Murphy} points out that the direction (1,1,0) manifests particularly strong two-dimensionality. 
 
We found that the c-axis static susceptibility increases under a field directed in the basal plane due to the proximity to the resonance and becomes larger than in the normal state. The wave-vector of the SDW is not constrained by the \emph{in-plane} orientation of the field (as remarked experimentally \cite{Kenzelmann}) but by the gapped energy spectrum characteristics. The conditions for stabilization of magnetism with respect to the normal state at finite magnetic field are the following: 
(i) The Fermi line must contain hotspots with ordering wave-vector close to $(1/2,1/2)$  corresponding to antiferromagnetic correlations). (ii) The two values of the gap on hot-spots must be non-zero and of opposite signs. As already emphasized\cite{Eremin} for CeCoIn$_5$, this requires the symmetry of the gap to be d$_{x^2-y^2}$-wave. This gives an upper boundary for the value of the field that induces SDW $\mu B_{SDW}<(\Delta_{hs1}+\Delta_{hs2})/2$, where $\mu=g\mu_B/2$ is the electron magnetic moment.

One way to treat at the same time superconducting and antiferromagnetic correlations was put forward by Scalapino\cite{Scalapino} who dealt with Coulomb's interaction in a Random Phase Approximation susceptibility $\chi(\mathbf{q},\omega)=\chi_0(\mathbf{q},\omega)/[1-U_q\chi_0(\mathbf{q},\omega)]$, where $\chi_0(\mathbf{q},\omega))$ is the susceptibility that includes superconducting correlations and $U_q$ is a momentum-dependent Hubbard Coulomb repulsion potential (this was originally introduced for q-independent interaction). In this model the conditions for a collective excitation (called spin exciton) to occur are $U_q\Re\mathfrak{e}\chi_0(\mathbf{q},\omega)=1$, and $U_q\Im\mathfrak{m}\chi_0(\mathbf{q},\omega)\ll1$.

Here superconductivity is accounted for in the mean field BCS theory for electrons under a magnetic field that couples to the electron spin via the Zeeman energy (the effect of vortices on the spin-exciton are discussed in\cite{Eschrig} in the context of high-T$_c$ superconductors). 
The Gor'kov Green's functions write once compacted into the Nambu (particle-hole space) notation 
$$\mathcal{G}_\sigma(i\omega_n,\mathbf{k})=\frac{(i\tilde{\omega}_n+\sigma \mu B)\tau_0+\Delta_k\tau_1+\xi_k\tau_3}{(i\tilde{\omega}_n+\sigma \mu B-E_k)(i\tilde{\omega}_n+\sigma \mu B+E_k)}.$$ 
Here  $\tilde{\omega}_n=\omega_n+\text{sign}(\omega_n)/(2\tau)$, $\omega_n=\pi T(n+1/2)$ are Matsubara frequencies, and $\tau$ is the electron relaxation time. We consider a gap order parameter $\Delta_k$ which is uniform in space and its phase was set equal to zero, $E_k=\sqrt{\xi_k^2+\Delta_k^2}$ is the zero-field energy of excitations in the superconducting state, $\tau_0$ is the unity matrix and $\tau_1$ and $\tau_3$ the Pauli matrices. Throughout we take the spin quantization axis the same as the magnetic field direction which is fixed to be the z-direction in our spin-space frame and belongs to the basal plane of the crystal.

The normal-state electron energy spectrum in CeCoIn$_5$ was calculated \cite{Tanaka} in a model involving hybridized conduction-electron bands and f-electron bands with Coulomb's interaction. 
Here we take the two-dimensional spectrum 
\begin{eqnarray}
\xi_k=2t[\cos(k_a)+\cos(k_b)]+4t'\cos(k_a)\cos(k_b)\nonumber\\
+2t''[\cos(2k_a)+\cos(2k_b)]+\varepsilon,
\end{eqnarray}
with $t'=-0.5t$, $t''=0.4t$ and $\varepsilon=0.6t$.
\begin{figure}[h!]
\centering
\includegraphics[width=6cm]{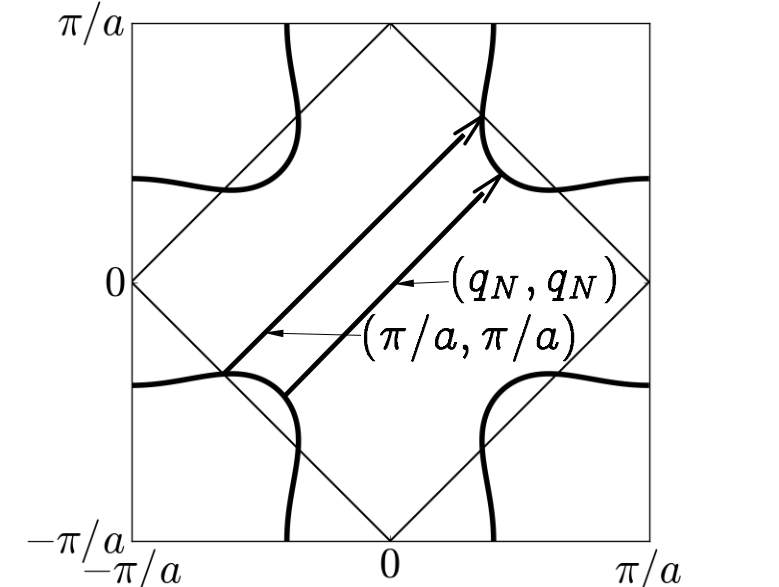}
\caption{Contour plot of the normal-state energy spectrum $\xi_k=0$ (the Fermi line) in the first Brillouin zone. The configurations of the commensurate and nesting wave-vectors are drawn.}
\label{Brillouin}
\end{figure}
We consider a singlet d$_{x^2-y^2}$-wave superconducting state with gap $\Delta_k=(\Delta_0/2)[\cos(k_a)-\cos(k_b)]$ corresponding to the irreducible representation $B_{1g}$ of the tetragonal crystal point group. In the following all energies are counted with respect to the scale given by the gap $\Delta_0$ at zero temperature and zero field, and we set $t/\Delta_0=7$.

The free electron spin susceptibility is defined as
\begin{eqnarray}\chi_0^{ab}(\mathbf{q},i\nu_m)~~~~~~~~~~~~~~~~~~~~~~~~~~~~~~~~~~\nonumber\\=-\frac{T}{2}\sum_{k,\omega_n,\sigma,\sigma'}\text{Tr}\left[\sigma_{\sigma\sigma'}^a\mathcal{G}_{\sigma'}(\mathbf{k},i\omega_n)\sigma_{\sigma'\sigma}^b\mathcal{G}_\sigma(\mathbf{k}+\mathbf{q},i\omega_n+i\nu_m)\right],\nonumber
\end{eqnarray} where the trace is performed in the particle-hole (Nambu) space and $\sigma_{\sigma\sigma'}^a$ are the Pauli matrices in spin space.
The susceptibility is calculated by taking account of the self-consistency equation for the order parameter, yielding slight decrease in the gap magnitude at high field. We evaluate the retarded susceptibility along the crystal c-axis (x-direction of the frame in spin-space introduced above) under a magnetic field applied in the crystal basal plane (z-direction) 
\begin{eqnarray}
 &&~~~~~~~~~~~~~~~~~~~~~~~~~~~~\chi_0^{xx}({\bf q},\omega)=\\
 &&\frac{1}{2}\sum_{\mathbf{k},\sigma} \Big\{-l^2(\mathbf{k},\mathbf{q})[f(E_k-\sigma\mu B)-f(E_{k+q}+\sigma\mu B)]\nonumber\\
 &&~~~~~~~~~~~~~~~\times\Big[\frac{1}{E_k-E_{k+q}-2\sigma\mu B+\omega +i/(2\tau)}\nonumber\\
 &&~~~~~~~~~~~~~~~~~~~~+\frac{1}{E_k-E_{k+q}-2\sigma\mu B-\omega-i/(2\tau)}\Big ]\nonumber\\
 &&~~~~~~~~+p^2(\mathbf{k},\mathbf{q})[1-f(E_k-\sigma\mu B)-f(E_{k+q}-\sigma\mu B)]\nonumber\\
 &&~~~~~~~~~~~~~~~\times\Big[\frac{1}{E_k+E_{k+q}-2\sigma\mu B+\omega+i/(2\tau)}\nonumber\\
 &&~~~~~~~~~~~~~~~~~~~~+\frac{1}{E_k+E_{k+q}-2\sigma\mu B-\omega-i/(2\tau)}\Big]\Big\}\nonumber.
\label{chi0}
\end{eqnarray}
Here  $\sum_{\mathbf{k}}=\int_{B.Z.}d^2ka^2/(2\pi)^2$, the integral being evaluated over the (first) Brillouin zone $k_a,k_b\in[-\pi/a,\pi/a]$. 
The \emph{coherence factors} are
\begin{eqnarray}
l^2(\mathbf{k},\mathbf{q})&=&\frac{1}{2}\left(1+\frac{\xi_k\xi_{k+q}+\Delta_k\Delta_{k+q}}{E_kE_{k+q}}\right),\\
p^2(\mathbf{k},\mathbf{q})&=&\frac{1}{2}\left(1-\frac{\xi_k\xi_{k+q}+\Delta_k\Delta_{k+q}}{E_kE_{k+q}}\right).
\end{eqnarray}
Under the condition of a superconducting gap having d$_{x^2-y^2}$-wave symmetry, the coherence factor $p^2$ is close to unity at the points of the Fermi line where $\Delta_k=-\Delta_{k+q}$. The importance of the symmetry of the order parameter was emphasized\cite{Lavagna,Fong} for the occurrence of the resonance. Throughout we considered the temperature $T=0.05\Delta_0$, the g-factor $g=2$, and the damping parameter $1/(2\tau)=0.02\Delta_0$.

\begin{figure}[h]
\centering
\includegraphics[width=7cm]{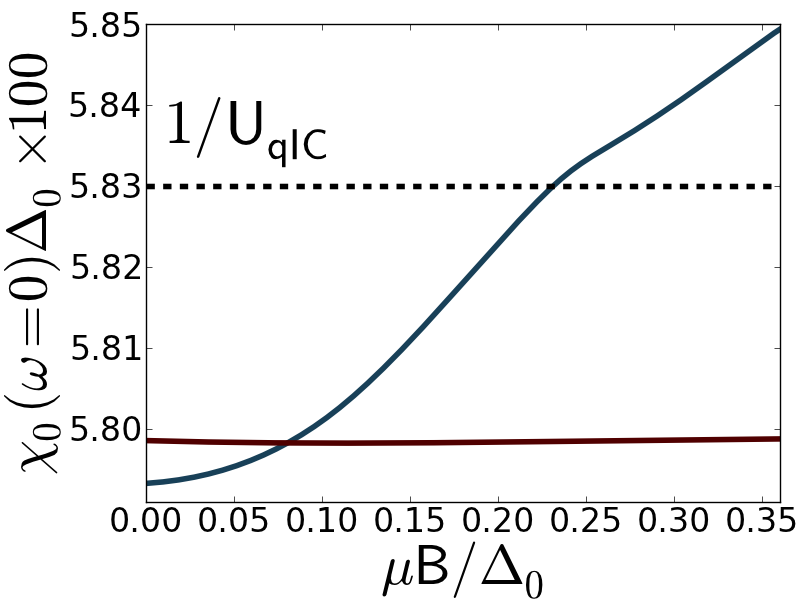}
\caption{Condition for obtaining an instability from the superconductivity to coexisting superconductivity and SDW ordering at finite field in the case of incommensurate wave-vector $q_{IC}=0.45$. The grey line represents the normal-state static susceptibility along the crystal c-axis $\chi_0^{xx}=\sum_{\mathbf{k}\sigma} [f(\xi_k-\sigma\mu B)-f(\xi_{k+q}+\sigma\mu B)]\times[\xi_k-\xi_{k+q}-2\sigma\mu B]/[(\xi_k-\xi_{k+q}-2\sigma\mu B)^2+1/(2\tau)^2]$ and the black line the superconducting one.}
\label{chi0}
\end{figure}

The field-dependence of the real static susceptibility $\chi_0^{xx}(w=0,B)$ is shown in Fig. (\ref{chi0}) for incommensurate ordering wave-vector and illustrates the condition for the magnetic instability to occur, which is $\chi_0^{xx}(\mathbf{q},\omega=0)=1/U_q$ with the additional condition for the existence of superconductivity $B<H_p$ (the zero temperature paramagnetic critical field was computed from the superconducting free energy with the band structure introduced above and was found to be $H_p=0.36\Delta_0/\mu$). Due to d$_{x^2-y^2}$-wave superconductivity the free susceptibility increases with field, becomes larger than the normal-state susceptibility and reaches a maximum at the value $B=\Delta_{hs}/\mu$. Before this point a field can possibly trigger a SDW instability in contrast to the normal-state case where $\chi_{0N}^{xx}(B)$ remains essentially constant. For arbitrary direction, the free spin susceptibility writes $\chi_0^{\varphi\varphi}=\cos^2(\varphi)\chi_0^{zz}+\sin^2(\varphi)\chi_0^{xx}$,
where $\varphi$ is the angle between the magnetic field and the c-axis. The longitudinal susceptibility $\chi_0^{zz}$ doesn't carry the field dependence in the denominator and therefore for $\mathbf{B}||c$ no magnetic ordering along the c-axis can be induced. This point is consistent with experiment\cite{Blackburn} where it was found that signal of magnetic ordering disappears as the angle between the field and the crystal plane increases.

\begin{figure}[h]
\centering
\includegraphics[width=7cm]{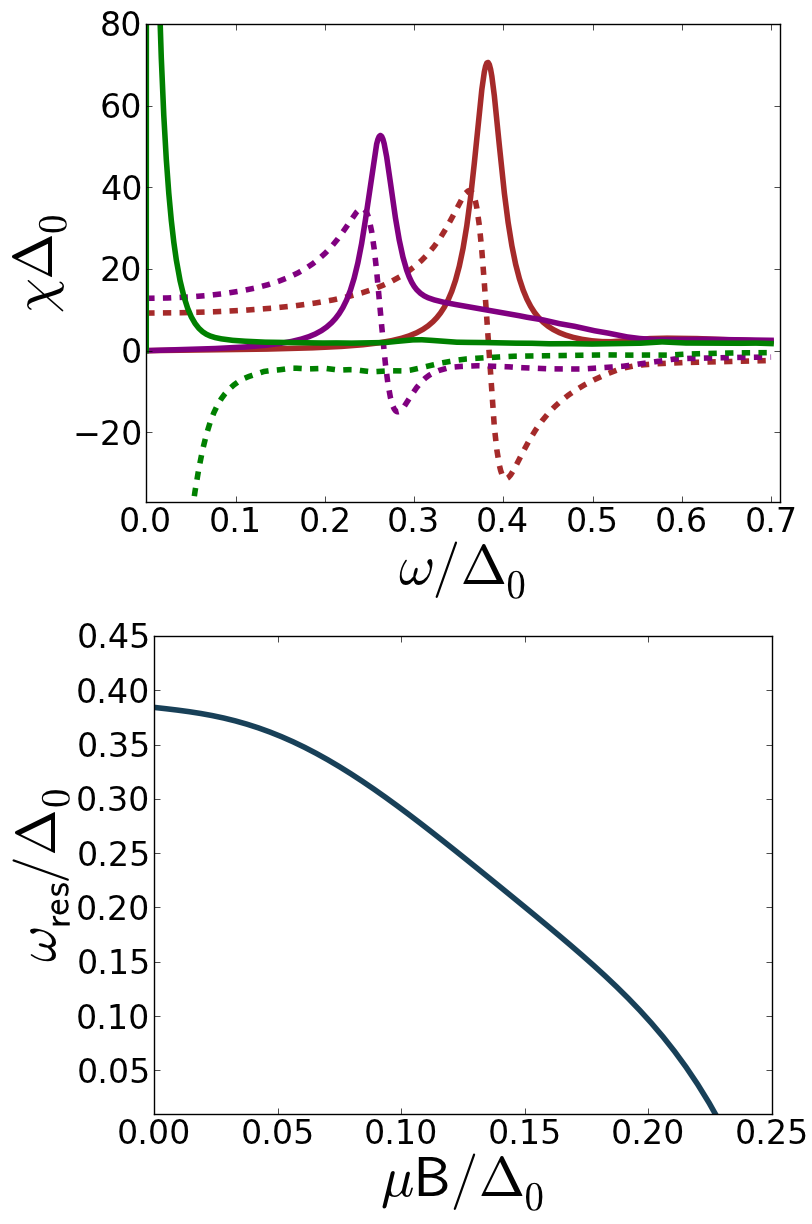}
\caption{Above: real (dotted lines) and imaginary (full lines) parts of the RPA susceptibility computed at magnetic fields $\mu B/\Delta_0=0$ (brown), $\mu B^\ast/(2\Delta_0)=0.11465$ (purple), and $\mu B^\ast/\Delta_0=0.2293$ (green). We used the potential energy value $U_{qIC}/\Delta_0=1/0.0583$. See text for discussion. Below: evolution of the resonance energy with a magnetic field applied parallel to the crystal plane. The value of the condensation field corresponding to vanishing resonance energy is found to be $B^\ast=0.2293\Delta_0/\mu$.}
\label{chi}
\end{figure}

The upper part of Fig. (\ref{chi}) shows the evolution with field of the collective excitation as given by the RPA dynamic susceptibility at wave-vector $\mathbf{Q}=\mathbf{Q}_{IC}=(0.45,0.45)$. The value of the gap at hot-spots is $\Delta_{hs}=0.27\Delta_0$. We here discuss several points: (i) The zero field collective peak appears at energy $\omega_{res}\simeq0.384\Delta_0\lesssim2\Delta_{hs}$. (ii) Under a transverse magnetic field a well defined collective peak is shifted to lower energy due to Zeeman splitting of the energy of elementary excitations. The higher energy feature of $\Im\mathfrak{m}\chi^{xx}(w)$ is not present since the excitations are strongly damped by the continuum. This observation is consistent with experiment\cite{Panarin}. (iii) The field at which the excitation softens to zero energy is $\mu B^\ast=0.2293\Delta_0\lesssim \omega_{res}/2$, hence showing the transition from the finite energy resonance (that corresponds to the condition $U_q\Re\mathfrak{e}\chi_0(\mathbf{q},\omega)=1$) to the ground-state instability at the spin-exciton condensation field. The evolution of the resonance energy with applied field is represented in the lower part of Fig. (\ref{chi}).  We emphasize here the necessity for the transition from superconducting to normal state to be of the first order since a vanishingly small gap would give negligibly small effect.
The physics behind the effect described here is reminiscent of the excitonic phases close to a semiconductor/semi-metal transition\cite{Halperin}. Here excitons originate from d$_{x^2-y^2}$-wave superconductivity under a time-dependent perturbation and magnetic ordering represents the zero-energy condensation of excitons at finite wave-vector under a transverse magnetic field.

CeCoIn$_5$ is a system close to antiferromagnetism and for this reason the interaction $U_q$ is considered to be maximum at the antiferromagnetic wave-vector and as a consequence the collective excitation dispersion is predicted\cite{Eremin} to be centered on the commensurate wave-vector. The dispersion shows\cite{Eremin} a downward shape and an incommensurate ground state is expected. This situation is however sensitive to the precise band structure, which is known\cite{Settai} to be quite complicated and to consist of several bands. Experimental determination in CeCoIn$_5$ of the excitation dispersion would represent progress on this issue. 

To summarize, we have presented a new approach for understanding magnetism that is tied with superconductivity in CeCoIn$_5$. The condensation of the induced by d$_{x^2-y^2}$-wave superconductivity spin-exciton driven by the Zeeman splitting of the energy of elementary excitations generates a state with coexisting magnetism and superconductivity.
There was found that the static RPA susceptibility along c-direction in quasi-2D-tetragonal d-wave superconductor under strong enough magnetic field applied in the basal plane  is singular.
The instability occurs at wave-vector $\mathbf{Q}$ that connects points of the Fermi line with a finite gap. Because of the proximity of the compound to antiferromagnetism this can be commensurate or incommensurate close to commensuration. This presents evidence that the feedback of d$_{x^2-y^2}$-wave superconductivity can have drastic consequences not only on collective excitations in the system but also on the ground state properties when a external parameter tunes the excitation energy to zero (here realised with magnetic field applied in the tetragonal crystal plane). Several questions on this transition however remain, in particular spectroscopic probe such as Electron Spin Resonance (free of low-energy incoherent peak) might help at studying closely the transition.

This work was partly supported by the grant SINUS of Agence Nationale de la Recherche. V. Michal would like to thank J. Panarin and S. Raymond for discussions on recent Inelastic Neutron Scattering experiments and also M. Lavagna for useful conversations.

\end{document}